\documentclass[aps,graphicx,12pt,showkeys,showpacs]{revtex4}
\usepackage{amsmath}
\usepackage{amscd}
\usepackage{graphicx}
\usepackage{subfigure}
\usepackage{appendix}

\begin{document}

\title{Fast generation of $N$-atom Greenberger-Horne-Zeilinger state in separate coupled cavities via transitionless quantum driving}

\author{Wu-Jiang Shan$^{1}$}
\author{Ye-Hong Chen$^{1}$}
\author{Yan Xia$^{1,}$\footnote{E-mail: xia-208@163.com}}
\author{Jie Song$^{2,}$\footnote{E-mail: jsong@hit.edu.cn}}

\affiliation{$^{1}$Department of Physics, Fuzhou University, Fuzhou
350002, China\\$^{2}$Department of Physics, Harbin Institute of
Technology, Harbin 150001, China}

%tableofcontents

\begin{abstract}
By jointly using quantum Zeno dynamics and the approach of
``transitionless quantum driving (TQD)" proposed by Berry to
construct shortcuts to adiabatic passage (STAP), we propose an
efficient scheme to fast generate multiatom
Greenberger-Horne-Zeilinger (GHZ) states in separate cavities
connected by opitical fibers only by one-step manipulation. We first
detail the generation of the three-atom GHZ states via TQD, then, we
compare the proposed TQD scheme with the traditional ones with
adiabatic passage. At last, the influence of various decoherence
factors, such as spontaneous emission, cavity decay and fiber photon
leakage, is discussed by numerical simulations. All the results show
that the present TQD scheme is fast and insensitive to atomic
spontaneous emission and fiber photon leakage. Furthermore, the
scheme can be directly generalized to realize $N$-atom GHZ states
generation by the same principle in theory.
\end{abstract}

\maketitle

\section{Introduction}
Quantum entanglement is not only one of the most important features
in quantum mechanics \cite{SBZGCGPRL}, but also a key resource for
testing quantum mechanics against local hidden theory
\cite{JSBPhys}. Recently, the entangled states have been applied in
many fields in quantum information processing (QIP), such as quantum
computing \cite{MANILC}, quantum cryptography \cite{AKEPRL}, quantum
teleportation \cite{YXJSPMLHSS,CHBGBCCRJAPWW}, quantum secret
sharing \cite{MHVBABPRL}, and so on. These promising applications
have greatly motivated the researches in the generation of entangled
states.

It is worth noting that a typical entangled state so-called
Greenberger-Horne-Zeilinger (GHZ) state
$|GHZ\rangle=\frac{1}{\sqrt{2}}(|000\rangle+|111\rangle)$ , first
proposed and named by Daniel M. Greenberger, Machael Horne and Anton
Zeilinger \cite{DMGMHASAZAJP}, has raised much interest. Contrary to
other entangled states, the GHZ state exhibits some special
features, such as it is the maximally entangled state and can
maximally violate the Bell inequalities \cite{SBZEPJD2009}. In 2001,
Zheng has proposed a scheme to test quantum mechanics against local
hidden theory without the Bell's inequalities by use of multiatom
GHZ state \cite{SBZPRL2001}. Therefore, great interest has arisen
regarding the significant role of GHZ state in the foundations of
quantum mechanics measurement theory and quantum communication. At
present, the first and main problem we face is how to generate GHZ
state by using current technologies. To our knowledge, in some
experimental systems, such as trapped ions systems \cite{DLEKSSJB},
photons systems \cite{ZZYCANZYYHBJWP,stjzxpprl07}, and atoms systems
\cite{JMRMBSHRmp01}, scientists have realized the generation of such
GHZ state. Recently, a promising experimental instrument named
cavity quantum electrodynamics (C-QED), which concerns the
interaction between the atom and the quantized field within cavity
\cite{ZCSYXJSHSSQIC}, has aroused much attention. Based on C-QED,
many theoretical schemes for generating GHZ state have been
proposed. For example, Li \emph{et al.} have proposed a scheme to
generate multiatom GHZ state under the resonant condition by Zeno
dynamics \cite{WALWLFOE}, but the scheme is sensitive to the atomic
spontaneous emission and fiber photonic leakage. Hao \emph{et al.}
have proposed an efficient scheme to generate mulitiatom GHZ state
under the resonant condition via adiabatic passage
\cite{SYHYXJSNBA}, but it takes too long time. Chen \emph{et al.}
have proposed a smart scheme to overcome the above drawbacks, but
the scheme needs to trap three atoms in one cavity
\cite{YHCYXQQCJSarXGHZ}, such design is difficult to manipulate each
atom in experiment and to construct a large-scale quantum network.

On the other hand, in modern quantum application field, an important
method to manipulate the states of a quantum system is adiabatic
passage, included ``rapid" adiabatic passage (RAP), stimulated Raman
adiabatic passage (STIRAP), and their variants \cite{XCILARDGOJGMPRL2010}. The adiabatic
passage covers the shortage with respect to errors or fluctuations
of the parameters compared with the resonant pulses, but its
evolution speed is very slow, so it may be useless in some cases. In
recent years, shortcuts to adiabatic passage (STAP), which
accelerates a slow adiabatic quantum process via a non-adiabatic
route, has aroused a great deal of attention. Many theoretical
proposals have been presented to realize QIP, such as fast
population transfer
\cite{XCJGMPRA2012,YHCYXQQCJSPRA,MLYXLTSJSNBAPRA,MLYXLTSJSLP}, fast entanglement
generation \cite{MLYXLTSJSNBAPRA,YHCYXQQCJSLPL}, fast implementation
of quantum phase gates \cite{YHCYXQQCJSPRA2015,YLQCWSLSXJSZPRA2015}, and so on. To our
knowledge, the main methods to construct effective shortcuts has two
forms: one is invariant-based inverse engineering based
Lewis-Riesenfeld invariant (IBLR) \cite{HRLWBRJMP} and the other is
transitionless quantum driving (TQD) \cite{Berry2009}, which is
pointed out by Berry. The two methods are strongly related \cite{XCETJGMPRA2011}, but also
have their own characteristics. For example, the former does not
need to modify the original Hamiltonian $H_0(t)$, but the algorithm
is suitable for some special physical models. The latter needs to
modify the original Hamiltonian $H_0(t)$ to the ``counter-diabatic
driving" (CDD) Hamiltonian $H(t)$ to speed up the quantum process.
The fixed Hamiltonian $H(t)$ can be obtained in theory, but it does
not usually exist in real experiment.

In addition, the quantum Zeno effect (QZE), first understood by Neumann
\cite{VMJ1932} and named by Misra and Sudarshan \cite{BMECGS1977},
 exhibits a especially experimental phenomenon that transitions between quantum states can be hindered by frequent measurement.
The system will evolve away from its initial state and remain in the so-called ``Zeno subspace" defined by the measure due to frequently projecting onto a multi-dimensional
subspace \cite{PFVGGMSPECGSPLA,PFSPASLSSPRA}. This is so-called quantum Zeno dynamics (QZD). Without making using of projection operators and non-unitary,
 ``a continuous coupling" can obtain the same quantum Zeno
effect instead of discontinuous measurements
\cite{PFSP2002,PFGMSP2009}. Now, we give a brief introduction of the
quantum Zeno dynamics in the form of continuous coupling
\cite{PFGMSP2009}. Suppose that the system and its continuously
coupling external system are governed by the total Hamiltonian
$H_{tot}=H_s+KH_e$, where $H_s$ is the Hamiltonian of the quantum
system to be investigated, $H_e$ is an additional Hamiltonian caused
by the interaction with the external system, $K$ is the coupling
constant. In the limit $K\rightarrow\infty$, the evolution operator
of system can be expressed as
$U(t)=\exp{[-it\sum_n(K\eta_nP_n+P_nH_sP_n)]}$, where $P_n$ is the
eigenprojection of $H_e$ corresponding to the eigenvalue $\eta_n$,
i.e., $H_eP_n=\eta_nP_n$ \cite{RCYGLTCZ}.

Inspired by the above useful works, we make use of Zeno dynamics and
TQD to construct STAP to generate $N$-atom GHZ state in C-QED. Our
scheme has the following advantages: (1) The atoms are trapped in
different cavities so that the single qubit manipulation is more
available in experiment. (2) The fast quantum entangled state
generation for multiparticle in spatially separated atoms can be
achieved in one step. (3) Numerical results show that our scheme is
not only fast, but also robust against variations in the
experimental parameters and decoherence caused by atomic spontaneous
emission and fiber photon leakage. In fact, further research shows
that, the total operation time for the scheme is irrelevant to the
number $N$ of qubits.

The paper is organized as follows. In section \textrm{II}, we give a
brief introduction to the approach of TQD proposed by Berry. In
section \textrm{III}, we introduce the physical modal and the
systematic approximation by QZD. In section \textrm{IV}, we propose
the schemes to generate the three-atom GHZ state via TQD and
adiabatic passage, respectively. The decoherence caused by various
factors is discussed by the numerical simulation. In section
\textrm{V}, we directly generalize the scheme in section \textrm{IV}
to generate $N$-atom GHZ states in one step. At last, we discuss the
experimental feasibility and make a conclusion about the scheme in
section \textrm{VI}.

\section{Transitionless quantum driving}
Suppose a system is dominated by a time-dependent Hamiltonian
$H_o(t)$ with instantaneous eigenvectors $|\phi_n(t)\rangle$ and
eigenvalues $E_n(t)$,
\begin{equation}
H_o(t)|\phi_n(t)\rangle=E_n(t)|\phi_n(t)\rangle.
\end{equation}
When a slow change satisfying the adiabatic condition \textbf{does},
the system governed by $H_o(t)$ can be expressed at time $t$
\begin{eqnarray}
|\psi(t)\rangle&=&e^{i\xi_n(t)}|\phi_n(t)\rangle, \cr\cr
\xi_n(t)&=&-\frac{1}{h}\int_0^tdt^\prime E_n(t^\prime)+i\int_0^tdt^\prime\langle\phi_n(t^\prime)|\partial_{t^\prime}\phi_n(t^\prime)\rangle,
\end{eqnarray}
where
$\partial_{t^\prime}=\frac{\partial}{\partial_{t^\prime}}$. Because
the instantaneous eigenstates $|\phi_n(t)\rangle$ do not meet the
Schr$\ddot{\texttt{o}}$dinger equation
$i\hbar\partial_t|\phi_n(t)\rangle=H_0|\phi_n(t)\rangle$, a finite
probability that the system is in the state $|\phi_{m\neq
n}(t)\rangle$ will occur during the whole evolution process even
under the adiabatic condition.

To construct the Hamiltonian $H(t)$ that drives the instantaneous eigenvector $|\phi_n(t)\rangle$ exactly, i.e., there are no transitions
between different eigenvectors during the whole evolution process, we define the unitary operator
\begin{eqnarray}
U=\sum_n{e^{i\xi_n(t)}|\phi_n(t)\rangle\langle\phi_n(0)|},
\end{eqnarray}
%it must obey
%\begin{equation}
%i\hbar\partial_tU=H(t)U,
%\end{equation}
we can formally solve the Schr\"{o}dinger equation
\begin{equation}
H(t)=i\hbar(\partial_tU)U^\dag.
\end{equation}
Substituting eq. (3) into eq.(4), the Hamiltonian $H(t)$ can be expressed
\begin{eqnarray}
H(t)&=&i\hbar\sum_n{(|\partial_t\phi_n\rangle\langle\phi_n|}-\hbar\sum_n{|\phi_n\rangle\dot{\xi_n}\langle\phi_n|)},
%&=&H_0(t)+H_1(t).
\end{eqnarray}
the simplest choice is $E_n=0$, for which the bare states $|\phi_n(t)\rangle$, with no phase factors, are driven by
\begin{eqnarray}
H(t)=i\hbar\sum_n{|\partial_t\phi_n\rangle\langle\phi_n|}.
\end{eqnarray}

\section{Physical modal and systematic approximation by QZD}

\begin{figure}
 \scalebox{1.0}{\includegraphics{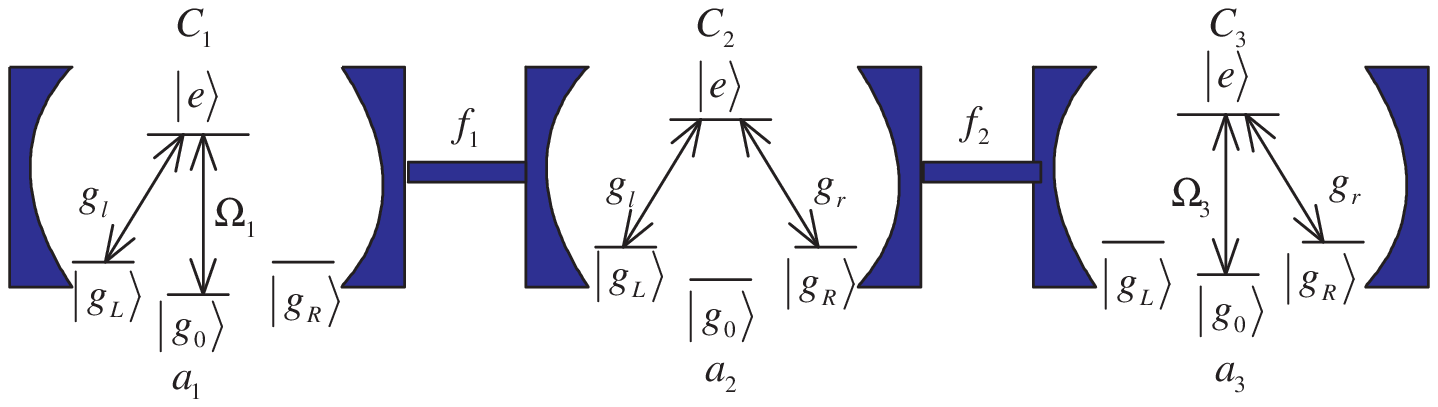}}
 \caption{The structure of the experimental setup and atoms. Three identical atoms
 $a_1$, $a_2$ and $a_3$ are trapped in three separated cavities
 $C_1$, $C_2$ and $C_3$, which are linked by two fibers $f_1$, $f_2$.}\label{FIG1}
\end{figure}

For the sake of the clearness, let us first consider the physical
modal that three identical atoms $a_1$, $a_2$ and $a_3$ are trapped
in three linearly arranged optical cavities $C_1$, $C_2$ and $C_3$,
respectively. As shown in FIG. \ref{FIG1}, each atom possesses one
excited level $|e\rangle$ and three ground states $|g_l\rangle$,
$|g_o\rangle$ and $|g_r\rangle$. The cavities $C_1$ and $C_3$ are
single-mode, the cavity $C_2$ are bi-mode. $C_1$, $C_2$ and $C_3$
are connected by the optical fibers $f_1$, $f_2$, respectively.
Assuming that the transition
$|e\rangle_{a_{1(3)}}\leftrightarrow|g_o\rangle_{a_{1(3)}}$ is
resonantly driven by a external classical field with the
time-dependent Rabi frequencie $\Omega_{1(3)}(t)$, while the
transition
$|e\rangle_{1(2)}\leftrightarrow|g_l\rangle_{1(2)}(|e\rangle_{2(3)}\leftrightarrow|g_r\rangle_{2(3)})$
is resonantly coupled to the left-circularly(right-circularly)
polarized cavity mode with the coupling constant $g_{l(r)}$,
respectively.

In the short-fiber limit, i.e., $(2L\bar{\nu})/(2\pi c)\ll 1$ ($L$
is the length of the fibers, $\bar{\nu}$ is the decay rate of the
cavity fields into a continuum of fiber modes and $c$ is
the speed of light), only one resonant mode of the fiber interacts
with the cavity mode \cite{ASSMSBPRL}. In the rotating frame, the
Hamiltonian of the whole system can be written as $(\hbar=1)$

\begin{eqnarray}\label{1}
H_{tot}&=&H_l+H_c,    \cr\cr
H_l&=&\sum_{o=1,3}\Omega_o(t)|e\rangle_{a_o}\langle g_0|+H.c.,  \cr\cr
H_c&=&g_{1l}a_{1l}|e\rangle_{a_1}\langle
g_l|+g_{2l}a_{2l}|e\rangle_{a_2}\langle
g_l|+g_{2r}a_{2r}|e\rangle_{a_2}\langle
g_{r}|+g_{3r}a_{3r}|e\rangle_{a_3}\langle g_r|\cr\cr
&&+v_1b_1^{\dag}(a_{1l}+a_{2l})+v_2b_2^{\dag}(a_{2r}+a_{3r})+H.c.,
\end{eqnarray}
where $a_{kl}^{\dag}\ (a_{kr}^{\dag})$ and $a_{kl}\ (a_{kr})$
 denote the creation and annihilation operators for the
left-circularly (right-circularly) polarized mode of cavities $C_k\
(k=1,\ 2,\ 3)$, respectively; $b_j^{\dag}$ and $b_j$ denote the
creation and annihilation operators associated with the resonant
mode of fiber $f_j\ (j=1,\ 2)$, respectively. For the sake of
simplicity, we assume $g_{1l}=g_{2l}=g_{2r}=g_{3r}=g$ and
$v_1=v_2=v$. If the initial state of the whole system is
$|g_og_lg_r\rangle|0\rangle_{C_1}|00\rangle_{C_2}|0\rangle_{C_3}|0\rangle_{f_1}|0\rangle_{f_2}$
(here $|g_og_lg_r\rangle = |g_og_lg_r\rangle_{a_1 a_2 a_3}$), the
whole system evolves in the following subspaces
\begin{eqnarray} |\phi_1\rangle&=&|g_og_lg_r\rangle|0\rangle_{C_1}|00\rangle_{C_2}|0\rangle_{C_3}|0\rangle_{f_1}|0\rangle_{f_2},
 ~|\phi_2\rangle=|eg_lg_r\rangle|0\rangle_{C_1}|00\rangle_{C_2}|0\rangle_{C_3}|0\rangle_{f_1}|0\rangle_{f_2},  \cr\cr
 |\phi_3\rangle&=&|g_lg_lg_r\rangle|1\rangle_{C_1}|00\rangle_{C_2}|0\rangle_{C_3}|0\rangle_{f_1}|0\rangle_{f_2},
 ~|\phi_4\rangle=|g_lg_lg_r\rangle|0\rangle_{C_1}|00\rangle_{C_2}|0\rangle_{C_3}|1\rangle_{f_1}|0\rangle_{f_2}, \cr\cr
 |\phi_5\rangle&=&|g_lg_lg_r\rangle|0\rangle_{C_1}|10\rangle_{C_2}|0\rangle_{C_3}|0\rangle_{f_1}|0\rangle_{f_2},
 ~|\phi_6\rangle=|g_leg_r\rangle|0\rangle_{C_1}|00\rangle_{C_2}|0\rangle_{C_3}|0\rangle_{f_1}|0\rangle_{f_2}, \cr\cr
 |\phi_7\rangle&=&|g_lg_rg_r\rangle|0\rangle_{C_1}|01\rangle_{C_2}|0\rangle_{C_3}|0\rangle_{f_1}|0\rangle_{f_2},
 ~|\phi_8\rangle=|g_lg_rg_r\rangle|0\rangle_{C_1}|00\rangle_{C_2}|0\rangle_{C_3}|0\rangle_{f_1}|1\rangle_{f_2}, \cr\cr
 |\phi_9\rangle&=&|g_lg_rg_r\rangle|0\rangle_{C_1}|00\rangle_{C_2}|1\rangle_{C_3}|0\rangle_{f_1}|0\rangle_{f_2},
 ~|\phi_{10}\rangle=|g_lg_re\rangle|0\rangle_{C_1}|00\rangle_{C_2}|0\rangle_{C_3}|0\rangle_{f_1}|0\rangle_{f_2}, \cr\cr
 |\phi_{11}\rangle&=&|g_lg_rg_o\rangle|0\rangle_{C_1}|00\rangle_{C_2}|0\rangle_{C_3}|0\rangle_{f_1}|0\rangle_{f_2},
\end{eqnarray}
where $|ijk\rangle\ (i,\ j,\ k\in[e,\ g_l,\ g_o,\ g_r])$ denotes the
state of the atoms in every cavity, $|n\rangle_s\ (s=C_1,\ C_3,\
f_1,\ f_2)$ means that the quantum field state of system contains
$n$ photons. $|n_1n_2\rangle_{C_2}$ means that the
number of left-circularly photon is $n_1$ and the number of
right-circularly photon is $n_2$ in the cavity $C_2$.

Under the Zeno condition $g,v\gg\Omega_1,\Omega_3$, the Hilbert subspace is split into nine invariant Zeno subspace
\begin{eqnarray}
Z_1&=&\{|\phi_1\rangle, |\psi_1\rangle, |\phi_{11}\rangle\},~Z_2=\{|\psi_2\rangle\}, \cr\cr
Z_3&=&\{|\psi_3\rangle\},~Z_4=\{|\psi_4\rangle\},~Z_5=\{|\psi_5\rangle\},~Z_6=\{|\psi_6\rangle\}, \cr\cr
Z_7&=&\{|\psi_7\rangle\},~Z_8=\{|\psi_8\rangle\},~Z_9=\{|\psi_9\rangle\},
\end{eqnarray}

where the eigenstates of $H_c$ are
\begin{eqnarray}
|\psi_1\rangle&=&N_1(|\phi_2\rangle-\frac{g}{v}|\phi_4\rangle+|\phi_6\rangle-\frac{g}{v}|\phi_8\rangle+|\phi_{10}\rangle), \cr\cr
 |\psi_2\rangle&=&N_2(-|\phi_2\rangle+\varepsilon_1|\phi_3\rangle-\eta_1|\phi_4\rangle-\chi_1|\phi_5\rangle+
\chi_1|\phi_7\rangle+\eta_1|\phi_8\rangle-\epsilon_1|\phi_9\rangle+|\phi_{10}\rangle), \cr\cr
|\psi_3\rangle&=&N_3(-|\phi_2\rangle-\varepsilon_1|\phi_3\rangle-\eta_1|\phi_4\rangle+\chi_1|\phi_5\rangle-
\chi_1|\phi_7\rangle+\eta_1|\phi_8\rangle+\epsilon_1|\phi_9\rangle+|\phi_{10}\rangle), \cr\cr
|\psi_4\rangle&=&N_4(|\phi_2\rangle-\mu_1|\phi_3\rangle-\zeta_1|\phi_4\rangle+\delta_1|\phi_5\rangle-\theta_1|\phi_6\rangle+
\delta_1|\phi_7\rangle-\zeta_1|\phi_8\rangle-\mu_1|\phi_9\rangle+|\phi_{10}\rangle), \cr\cr
|\psi_5\rangle&=&N_5(|\phi_2\rangle+\mu_1|\phi_3\rangle-\zeta_1|\phi_4\rangle-\delta_1|\phi_5\rangle-\theta_1|\phi_6\rangle-
\delta_1|\phi_7\rangle-\zeta_1|\phi_8\rangle+\mu_1|\phi_9\rangle+|\phi_{10}\rangle), \cr\cr
|\psi_6\rangle&=&N_6(-|\phi_2\rangle+\varepsilon_2|\phi_3\rangle-\eta_2|\psi_4\rangle+\chi_2|\phi_5\rangle-\chi_2|\phi_7\rangle+
\eta_2|\phi_8\rangle-\epsilon_2|\phi_9\rangle+|\phi_{10}\rangle), \cr\cr
|\psi_7\rangle&=&N_7(-|\phi_2\rangle-\varepsilon_2|\phi_3\rangle-\eta_2|\psi_4\rangle-\chi_2|\phi_5\rangle+\chi_2|\phi_7\rangle+
\eta_2|\phi_8\rangle+\epsilon_2|\phi_9\rangle+|\phi_{10}\rangle), \cr\cr
|\psi_8\rangle&=&N_8(|\phi_2\rangle-\mu_2|\phi_3\rangle+\zeta_2|\phi_4\rangle-\delta_2|\phi_5\rangle+\theta_2|\phi_6\rangle-
\delta_2|\phi_7\rangle+\zeta_2|\phi_8\rangle-\mu_2|\phi_9\rangle+|\phi_{10}\rangle), \cr\cr
|\psi_9\rangle&=&N_9(|\phi_2\rangle+\mu_2|\phi_3\rangle+\zeta_2|\phi_4\rangle+\delta_2|\phi_5\rangle+\theta_2|\phi_6\rangle+
\delta_2|\phi_7\rangle+\zeta_2|\phi_8\rangle+\mu_2|\phi_9\rangle+|\phi_{10}\rangle),
\end{eqnarray}
with the corresponding eigenvalues
\begin{eqnarray} \lambda_1&=&0,  ~\lambda_2=-\sqrt{(g^2+2v^2-A)/2},  ~\lambda_3=\sqrt{(g^2+2v^2-A)/2},   \cr\cr
\lambda_4&=&-\sqrt{(3g^2+2v^2-A)/2},  ~\lambda_5=\sqrt{(3g^2+2v^2-A)/2},  ~\lambda_6=-\sqrt{(g^2+2v^2+A)/2},  \cr\cr
\lambda_7&=&\sqrt{(g^2+2v^2+A)/2},  ~\lambda_8=-\sqrt{(3g^2+2v^2+A)/2},  ~\lambda_9=\sqrt{(3g^2+2v^2+A)/2},
\end{eqnarray}
where the parameters are
\begin{eqnarray}
\epsilon_1&=&\frac{\sqrt{g^2+2v^2-A}}{\sqrt{2}g},~\eta_1=\frac{-g^2+2v^2-A}{2gv},~\chi_1=\frac{\sqrt{g^2+2v^2-A}(g^2+A)}{2\sqrt{2}gv^2},\cr\cr
\mu_1&=&\frac{\sqrt{3g^2+2v^2-A}}{\sqrt{2}g},~\zeta_1=\frac{-g^2-2v^2+A}{2gv},~\delta_1=\frac{\sqrt{3g^2+2v^2-A}(-g^2+A)}{2\sqrt{2}gv^2},~\theta_1=\frac{-g^2+A}{v^2},\cr\cr
\epsilon_2&=&\frac{\sqrt{g^2+2v^2+A}}{\sqrt{2}g},~\eta_2=\frac{-g^2+2v^2+A}{2gv},~\chi_2=\frac{\sqrt{g^2+2v^2+A}(-g^2+A)}{2\sqrt{2}gv^2},\cr\cr
\mu_2&=&\frac{\sqrt{3g^2+2v^2+A}}{\sqrt{2}g},~\zeta_2=\frac{g^2+2v^2+A}{2gv},~\delta_2=\frac{\sqrt{3g^2+2v^2+A}(g^2+A)}{2\sqrt{2}gv^2},~\theta_2=\frac{g^2+A}{v^2},
\end{eqnarray}
in addition, $A=\sqrt{g^4+4v^4}$ and $N_w$ is the normalization
factor of the eigenstate $|\psi_w\rangle\ (w=1,2,\cdots,9)$.

The projector in the $k$th Zeno subspace $Z_k$ is
\begin{eqnarray}
P_k^\beta=|\beta\rangle\langle\beta|,\   (|\beta\rangle\in Z_k).
\end{eqnarray}
The Hamiltonian in Eq. (8) can be approximately given by
\begin{eqnarray}
H_{tot}&\simeq&\sum_{k,\beta,\gamma}{\lambda_kP_k^\beta+P_k^\beta H_lP_k^\gamma} \cr\cr
&=&\sum_{k=2}^9{\lambda_k|\psi_k\rangle\langle\psi_k|+N_1(\Omega_1|\phi_1\rangle\langle\psi_1|+\Omega_3|\phi_{11}\rangle\langle\psi_1|+H.c.)}.
\end{eqnarray}
If the initial state is $|g_og_lg_r\rangle|0\rangle_{C_1}|00\rangle_{C_2}|0\rangle_{C_3}|0\rangle_{f_1}|0\rangle_{f_2}$,
it reduces to
\begin{eqnarray}
H_{eff}=N_1(\Omega_1|\phi_1\rangle\langle\psi_1|+\Omega_3|\phi_{11}\rangle\langle\psi_1|+H.c.),
\end{eqnarray}
which can be treated as a simple three-level system with an excited
state $|\psi_1\rangle$ and two ground states $|\phi_1\rangle$ and
$|\phi_{11}\rangle$. Then we obtain the eigenvectors and eigenvalues
of the effective Hamiltonian $H_{eff}$ as
\begin{eqnarray}\label{eqdd}
|\eta_o(t)\rangle=\left(
\begin{array}{c}
      \cos\theta(t) \\
      0\\
      -\sin\theta(t)
\end{array}
\right),
~|\eta_{\pm}(t)\rangle=\frac{1}{\sqrt{2}}\left(
\begin{array}{c}
      \sin\theta(t) \\
      \pm1\\
      \cos\theta(t)
\end{array}
\right),
\end{eqnarray}
with the corresponding eigenvalues $\eta_0=0$ and $\eta_{\pm}=\pm
N_1\Omega$, and $\tan\theta=\frac{\Omega_1}{\Omega_3}$ and
$\Omega=\sqrt{\Omega_1^2+\Omega_3^2}$.

\section{The generation of the three-atom GHZ state via transitionless quantum driving and adiabatic passage}

\subsection{Adiabatic passage method}

For the sake of the clearness, we first briefly present how to
generate the three-atom GHZ state via adiabatic passage.
When the adiabatic condition $|\langle n_0|\partial_tn_\pm\rangle|\ll|\lambda_{\pm}^\prime|$ is fulfilled well
and the initial state is $|\psi(0)\rangle=|\phi_1\rangle$, the state evolution will always follow
$|n_0(t)\rangle$ closely. To generate the three-atom GHZ states via the adiabatic passage and meet the boundary
conditions of the fractional stimulated Raman adiabatic passage (STIRAP),
\begin{eqnarray}
\lim_{t\rightarrow-\infty}{\frac{\Omega_1(t)}{\Omega_3(t)}}=0,
~\lim_{t\rightarrow+\infty}{\frac{\Omega_3(t)}{\Omega_1(t)}}=\tan{\alpha},
\end{eqnarray}
we need properly to tailor the Rabi frequencies $\Omega_1(t)$ and $\Omega_3(t)$ in the original
Hamiltonian $H_{tot}$
\begin{eqnarray}
\Omega_1(t)&=&\sin{\alpha}\Omega_0\exp{[\frac{-(t-t_0-t_f/2)^2}{t_c^2}]}, \cr\cr
\Omega_3(t)&=&\Omega_0\exp{[\frac{-(t+t_0-t_f/2)^2}{t_c^2}]}+\cos{\alpha}\Omega_0\exp{[\frac{-(t-t_0-t_f/2)^2}{t_c^2}]},
\end{eqnarray}
where $\Omega_0$ is the pulse amplitude and $t_f$ is the operation
time. $t_c$ and $t_0$ are some related parameters to be chosen for
the best performance of the adiabatic passage process. In order to
achieve better performance and meet the boundary conditions, we
suitably chose the parameters that $\tan{\alpha}=1$, $t_0=0.14t_f$
and $t_c=0.19t_f$. As shown in Fig. 2, the time-dependent
$\Omega_1(t)/\Omega_0$ and $\Omega_3(t)/\Omega_0$ versus $t/t_f$ are
plotted with the fixed values $t_0$ and $t_c$. With the above
parameters, we obtain our wanted three-atom GHZ state
$|\psi(t_f)\rangle=(|\phi_1\rangle-|\phi_{11}\rangle)/\sqrt{2}$ via
the adiabatic passage. But this evolution process needs a relatively
long time to satisfy the adiabatic condition. We will detail the
reasons in the section of numerical simulations and analyses.

\subsection{Transitionless quantum driving method}

To reduce the evolution time and obtain the same state as the
adiabatic passage, we use the approach of TQD to construct STAP. As
introduced in the above, STAP speeds up a slow adiabatic passage via
a non-adiabatic passage route to achieve a same outcome, and the TQD
method is a important route to construct shortcuts. According to the
ideas proposed by Berry \cite{Berry2009}, the instantaneous states
in Eq. (16) do not meet the Schr\"{o}dinger equation, i.e.,
$i\partial_t|n_k\rangle\neq H_{eff}|n_k\rangle(k=0,\pm)$, so the
situation that the system starts from the state $|\psi_n(0)\rangle$
and ends up in the state $|\psi_{m\neq n}(t)\rangle$ occurs in a
finite probability even under the adiabatic condition. To drive the
instantaneous states $|n_k\rangle(k=0,\pm)$ exactly, we look for a
Hamiltonian $H(t)$ related to the original Hamiltonian $H_{eff}$
according to Berry's transitionless tracking algorithm
\cite{Berry2009}. From section $\textrm{II}$, we know the simplest
Hamiltonian $H(t)$ possessed the form,
\begin{eqnarray}
H(t)=i\sum_{0,\pm}|\partial_tn_k(t)\rangle\langle n_k(t)|.
\end{eqnarray}
Substituting Eq. (16) in Eq. (19), we obtain
\begin{eqnarray}
H(t)=i\dot{\theta}|\phi_1\rangle\langle\phi_{11}|+H.c.,
\end{eqnarray}
where $\dot{\theta}=[\dot{\Omega}_1(t)\Omega_3(t)-\Omega_1(t)\dot{\Omega}_3(t)]/\Omega^2$.
This is our wanted CCD Hamiltonian to construct STAP, and we will detail how to construct
this Hamiltonian in experiment later.

For the present system, the CDD Hamiltonian $H(t)$ is given in Eq.
(18), but it is irrealizable under current experimental condition.
Inspired by Refs. [17, 19], we find an alternative physically
feasible (APF) Hamiltonian whose effect is equivalent to $H_1(t)$.
The design is shown in Fig. 3, the atomic transitions is not
resonantly coupled to the classical lasers and cavity modes with the
detuning $\Delta$. The Hamiltonian of the system reads
$H_{tot}^\prime=H_c+H_l+H_d$, where
$H_d=\sum_{k=1}^3{\Delta|e\rangle_k\langle e|}$. Then, similar to
the approximation by QZD in section \textrm{III}, we also obtain an
effective Hamiltonian for the non-resonant system
\begin{eqnarray}
H_{eff}^\prime=N_1(\overline{\Omega}_1|\phi_1\rangle\langle\psi_1|+\overline{\Omega}_3|\phi_{11}\rangle\langle\psi_1|+H.c.)
+3\Delta N_1^2|\psi_1\rangle\langle\psi_1|.
\end{eqnarray}
When the large detuning condition $3\Delta N_1\geq\overline{\Omega}_1,\ \overline{\Omega}_3$ is satisfied,
we can adiabatically eliminate the state
$|\psi_1\rangle$ and obtain the final effective Hamiltonian
\begin{eqnarray}\label{eq22}
H_{fe}=-\frac{\overline{\Omega}_1^2}{3\Delta}|\phi_1\rangle\langle\phi_1|-\frac{\overline{\Omega}_3^2}{3\Delta}|\phi_{11}\rangle\langle\phi_{11}|
-\frac{\overline{\Omega}_1\overline{\Omega}_3}{3\Delta}(|\phi_1\rangle\langle\phi_{11}|+|\phi_{11}\rangle\langle\phi_1|).
\end{eqnarray}
For simplicity, we set $\overline{\Omega}_1=\overline{\Omega}_3=\overline{\Omega}(t)$.
The front two terms caused by Stark shift can be removed and the Hamiltonian becomes
\begin{eqnarray}\label{eq23}
\overline{H}(t)=\Omega_x|\phi_1\rangle\langle\phi_{11}|+H.c.,
\end{eqnarray}
where $\Omega_x=-\frac{\overline{\Omega}^2}{3\Delta}$. The equation
has a similar form with Eq. (20), but the effective couplings
between $i\dot{\theta}$ and $\Omega_x$ exist $3\pi/2$-dephased. To
guarantee their consistency, we put a change that
$\Omega_3\rightarrow-i\Omega_3$. Then, the eigenstates of $H_{eff}$
become
\begin{eqnarray}
|\eta_o^\prime(t)\rangle=\left(
\begin{array}{c}
      \cos\theta(t) \\
      0\\
      i\sin\theta(t)
\end{array}
\right),
~|\eta_{\pm}^\prime(t)\rangle=\frac{1}{\sqrt{2}}\left(
\begin{array}{c}
      \sin\theta(t) \\
      \pm1\\
      -i\cos\theta(t)
\end{array}
\right),
\end{eqnarray}
and the CDD Hamiltonian $H(t)$ becomes
\begin{eqnarray}
H(t)=-\dot{\theta}|\phi_1\rangle\langle\phi_{11}|-\dot{\theta}|\phi_{11}\rangle\langle\phi_1|.
\end{eqnarray}
Compared Eq. (23) with Eq. (25), we can easily get the CDD Hamiltonian
when the condition $\Omega_x=-\dot{\theta}$ is satisfied.
\begin{eqnarray}
\overline{\Omega}_1(t)=\overline{\Omega}_3(t)=\overline{\Omega}(t)=\sqrt{3\Delta\dot{\theta}}.
\end{eqnarray}

\subsection{Numerical simulations and analyses}

Next we will show that it takes less time to get the target state on
the situation governed by the APF Hamiltonian $H_{tot}^\prime$ via
TQD than by the original Hamiltonian $H_{tot}$ via adiabatic
passage. The time-dependent population for any state $|\psi\rangle$
is defined as $P=|\langle\psi|\rho(t)|\psi\rangle|$, where $\rho(t)$
is the corresponding time-dependent density operator. We present the
fidelity versus the laser pulses amplitude $\Omega_o$ and the
operation time $t/t_f$ via adiabatic passage in Fig. 4. As shown in
Fig. 4, we can know that the bigger the laser pulse amplitude is,
the less time that the system evolution to the target state needs.
However, we need to satisfy the Zeno conditions
$g,v\gg\Omega_1,\Omega_3$, so we set $\Omega_0=0.2g$. In Fig. 5, we
display the time-dependent populations of the states
$|\phi_1\rangle$, $|\psi_{target}\rangle$, and $|\phi_{11}\rangle$
via adiabatic passage. As depicted in Fig. 4 and Figs. 5, the
operation time needs $t_f\geq400/g$ to achieve an ideal result at
least. It is awkward in some case.

Next we will detail the evolution governed by the APF Hamiltonian
$H_{tot}^\prime$ via TQD. According to eq. (24) we finally get a GHZ
state
$|\psi(t_f)=\frac{1}{\sqrt{2}}(|\phi_1\rangle+i|\phi_{11}\rangle)$.
In Fig. 6, we present the relationship between the fidelity of the
three-atom GHZ state (governed by the APF Hamiltonian) and two
parameters $\Delta$ and $t_f$ when $\Omega_0=0.2g$ to satisfy the
Zeno condition, where the fidelity of the three-atom GHZ state is
defined as $F=|\langle GHZ|\rho(t_f)|GHZ\rangle|$ ($\rho(t_f)$ is
the desity operator of the whole system when $t=t_f$). We find that
a wide range for parameters $\Delta$ and $t_f$ can obtain a high
fidelity of the three-atom GHZ state, and the fidelity increases
with the increasing of $\Delta$ and the decreasing of $t_f$. In
order to satisfy the large detuning condition, we set $\Delta=2.3g$.
The Fig. 6 reveals that the operation time needs $t_f\geq72/g$ via
TQD at least.
%This result can be understood from putting Eq. () into Eq. (), we obtain a relationship
%\begin{eqnarray}
%\overline{\Omega}_0\propto\sqrt{\frac{\Delta}{t_f}}
%\end{eqnarray}
%where $\overline{\Omega}_0$ is the amplitude of $\overline{\Omega}$. In order to meet the Zeno condition
%$\overline{\Omega}\ll g,\ v$ and the large detuning condition $\overline{\Omega}\ll 3\Delta N_1$, the ratio should be small enough.
In Figs. 7 we plot the operation time for the creation of the GHZ state governed by $H_{tot}^\prime$ and
by $H_{tot}$ with the parameters that $t_f=72/g$, $\Omega_0=0.2g$, $\Delta=2.3g$ and $g=v$.
Numerical results show that the APF Hamiltonian $H_{tot}^\prime$ can
govern the evolution to a perfect GHZ state $|\psi(t_f)\rangle$ from $|\psi_1\rangle$ in a relatively short
interaction time while the original Hamiltonian $H_{tot}$ can not.

In above analysis, we do not consider the influence of decoherence caused by various factors, such as
spontaneous emissions, cavity decays and fiber photon leakages. In fact, the decoherence is unavoidable
during the evolution of the whole system in experiment. The master equation of the whole system is written as
\begin{eqnarray}
\dot{\rho}&=&-i [H_{tot},\rho] \cr\cr
           &&+\sum_{k=1}^3\frac{\gamma_k}{2}(2\sigma_k^-\rho\sigma_k^+-\sigma_k^+\sigma_k^-\rho-\rho\sigma_k^+\sigma_k^-) \cr\cr
           &&+\sum_{k=1}^2\frac{\kappa_{c_k}}{2}(2a_{l,k}\rho a_{l,k}^+-a_{l,k}^+a_{l,k}\rho-\rho a_{l,k}^+a_{l,k}) \cr\cr
           &&+\sum_{k=2}^3\frac{\kappa_{c_k}}{2}(2a_{r,k}\rho a_{r,k}^+-a_{r,k}^+a_{r,k}\rho-\rho a_{r,k}^+a_{r,k}) \cr\cr
           &&+\sum_{k=1}^2\frac{\kappa_{f_k}}{2}(2b_k\rho b_k^+-b_k^+b_k\rho-\rho b_k^+b_k),
\end{eqnarray}
where $\gamma_k$ is the atomic spontaneous emission rate for the
$k$th atom and $\kappa_{c(f)}$ is the decay rate of the $k$th cavity
($k$th fiber), $\sigma_k^-$ denotes the atomic transition from the
ground states $|m\rangle\ (m=g_0,\ g_l,\ g_r)$ to the excited state
$|e\rangle$. For the sake of simplicity, we assume that
$\gamma_1=\gamma_2=\gamma_3=\gamma$,
$\kappa_{c_1}=\kappa_{c_2}=\kappa_{c_3}=\kappa_c$ and
$\kappa_{f_1}=\kappa_{f_2}=\kappa_f$. As shown in Fig. 8, we plot
the fidelity governed by the APF Hamiltonian $H_{tot}^\prime$ and by
the original Hamiltonian $H_{tot}$ and the dimensionless parameters
$\gamma/g$, $\kappa_c/g$ and $\kappa_f/g$, respectively. We can draw
a conclusion that the fidelities are almost unaffected by the fiber
decay both via TQD and via adiabatic passage. We focus on the main
decoherence factors included the cavity decay and the atomic
spontaneous emission. As shown in Figs. 9, we plot the fidelity
versus the cavity decay and the atomic spontaneous emission. We can
know the most important decoherence factor is the cavity decay. This
result can be understood from Ref. [18] that if the Zeno condition
can not be satisfied very well, the populations of the intermediate
states including the cavity excited states can not be suppressed
ideally.

From the above anslysis, we can obviously know that the evolution
time from the initial state to the target state via TQD is
$t_f=72/g$ when $\Omega_0=0.2g$, $\Delta=2.3g$, $t_0=0.14t_f$,
$t_c=0.19t_f$ and $g=v$, while the evolution time for the adiabatic
passage is $t_f=400/g$ when $\Omega_0=0.2g$, $t_0=0.14t_f$,
$t_c=0.19t_f$ and $g=v$. So, the benefit of the TQD method is shown
obviously that the speed via TQD method is faster than that via
adiabatic passage. It is more worthy to  note that the fidelity of
the target state via TQD is almost equal to that via adiabatic
passage. So our scheme has a huge advantage compared with the
proposals via adiabatic passage. That means the present scheme via
STAP method is not only fast but also robust.

As we all know, it is necessary for a good scheme to tolerate the
deviations of the experimental parameters, because it is impossible
to avoid the operational imperfection in experiment. Define that
$\delta x=x^{\prime}-x$ is the deviation of the ideal value $x$,
$x^{\prime}$ is the actual value. In Fig. 10, we plot the fidelity
of the target state $|\psi_{target}\rangle$ versus the deviations of
the experimental parameters $g$, $v$, $\Omega_0$, and $T$ ($T=t_f$
denotes the operation time). Numerical results demonstrate that our
scheme is robust against the fluctuation of the experimental
parameters.

\section{the generation of the $N$-atom GHZ state via transitionless quantum driving}

Next we briefly present the generalization of the scheme in Section
\textrm{IV} to generate $N$-atom GHZ state by the same principle. We
consider the physical configuration shown in Fig. 11, where $N$
atoms $a_1,\ a_2,\ \cdots,\ a_N$ are trapped in $N$ cavities $C_1,\
C_2,\ \cdots,\ C_N$ connected by $N-1$ fibers $f_1,\ f_2,\ \cdots,\
f_{N-1}$, respectively. The level configurations of the atoms
between two ends are the same as that of the atom $a_2$ in the
three-atom case, and the level configurations of $a_1$ and $a_N$ are
the same as those of $a_1$ and $a_3$ in the three-atom case,
respectively. The Hamiltonian of the present system can be written
as in the rotation framework
\begin{eqnarray}
H_{total}&=&H_l^{\prime}+H_{c}^{\prime}, \cr\cr
H_l^{\prime}&=&\Omega_1^\prime|e\rangle_{a_1}\langle g_0|
+\Omega_N^\prime|e\rangle_{a_N}\langle g_0|+H.c.,\cr\cr
H_{c}^{\prime}&=&\sum_{i=1}^{N-1}{g_{i,l}a_{i,l}|e\rangle_{a_i}\langle
g_l|}+\sum_{j=2}^{N}{g_{j,r}a_{j,r}|e\rangle_{a_j}\langle g_r|}
\cr\cr
&+&\sum_{k=2}^{N-1}{[v_{k-1}b_{k-1}^{\dag}(a_{k-1,l}+a_{k,l})+v_{k}b_{k}^{\dag}(a_{k,r}+a_{k+1,r})]}+H.c..
\end{eqnarray}
Let us consider the situation where $N$ is an odd number, i.e.,
$N=2l+1,\ (l=1,\ 2,\ 3, \cdots)$. Suppose that the initial state of
the atoms is $|g_0g_lg_rg_lg_r\ \cdots\ g_r\rangle$ while all the
cavities and fibers are vacuum, then the system can be expended in
the following subspace
\begin{eqnarray}
&&|\phi_1^{\prime}\rangle=|g_0g_lg_r\ \cdots\
g_r\rangle|0\rangle_{all},  ~|\phi_2^{\prime}\rangle=|eg_lg_r\
\cdots\ g_r\rangle|0\rangle_{all},\cr\cr
&&|\phi_3^{\prime}\rangle=|g_lg_lg_r\ \cdots\
g_r\rangle|1\rangle_{c_1},  ~|\phi_4^{\prime}\rangle=|g_lg_lg_r\
\cdots\ g_r\rangle|1\rangle_{f_1}, \cr\cr
&&|\phi_5^{\prime}\rangle=|g_lg_lg_r\ \cdots\
g_r\rangle|10\rangle_{c_2},  ~|\phi_6^{\prime}\rangle=|g_leg_r\
\cdots\ g_r\rangle|0\rangle_{all},  \cr\cr
&&|\phi_7^{\prime}\rangle=|g_lg_rg_r\ \cdots\
g_r\rangle|01\rangle_{c_2},  ~|\phi_8^{\prime}\rangle=|g_lg_rg_r\
\cdots\ g_r\rangle|1\rangle_{f_2},  \cr\cr
&&|\phi_9^{\prime}\rangle=|g_lg_rg_r\ \cdots\
g_r\rangle|01\rangle_{c_3},  ~|\phi_{10}^{\prime}\rangle=|g_lg_re\
\cdots\ g_r\rangle|0\rangle_{all},  \cdots \cr\cr
&&|\phi_{4N-1}^{\prime}\rangle=|g_lg_r\ \cdots\
g_rg_0\rangle|0\rangle_{all},
\end{eqnarray}
where $|0\rangle_{all}$ means that there is none photon in all boson
modes, $|n_1 n_2\rangle_{s_i}\ (s=C,\ f.\ i=1,\ 2, \cdots, N)$ means
that there are $n_1$ left-circularly photon and $n_2$
right-circularly photon in the corresponding cavity $C_i$ or fiber
$f_i$.

Similar to the above procedure from Eq. (10) to Eq. (16), we get an effective Hamiltonian
\begin{eqnarray}
H_{eff(N)}=N_1^{\prime}(\Omega_1^\prime|\phi_1^\prime\rangle\langle\psi_1^\prime|+
\Omega_N^\prime|\phi_{4N-1}^\prime\rangle\langle\psi_1^\prime|+H.c.),
\end{eqnarray}
where
\begin{eqnarray}
|\psi_1^\prime\rangle=N_1^\prime(\sum_{i=1}^N|\phi_{4i-2}^\prime\rangle-
\sum_{i=1}^{N-1}\frac{g}{v}|\phi_{4i}^\prime\rangle).
\end{eqnarray}
In addition, the eigenstates and eigenvalues of the Hamiltonian in Eq. (30) can be written as
\begin{eqnarray}
|\chi_o(t)\rangle=\left(
\begin{array}{c}
      \cos\theta^{\prime}(t) \\
      0\\
      -\sin\theta^{\prime}(t)
\end{array}
\right),
~|\chi_{\pm}(t)\rangle=\frac{1}{\sqrt{2}}\left(
\begin{array}{c}
      \sin\theta^{\prime}(t) \\
      \pm1\\
      \cos\theta^{\prime}(t)
\end{array}
\right),
\end{eqnarray}
with the corresponding eigenvalues $\chi_0^\prime=0$ and
$\chi_{\pm}^\prime=\pm N_1^{\prime}\Omega^{\prime}$, where
$\tan\theta^{\prime}=\frac{\Omega_1^{\prime}}{\Omega_N^{\prime}}$ and
$\Omega^{\prime}=\sqrt{\Omega_1^{\prime2}+\Omega_N^{\prime2}}$.
Substituting Eq. (32) in Eq. (19), we obtain
\begin{eqnarray}
H^{\prime}(t)=i\dot{\theta}^{\prime}|\phi_1^{\prime}\rangle\langle\phi_{4N-1}^{\prime}|+H.c.,
\end{eqnarray}
where $\dot{\theta}^{\prime}=[\dot{\Omega}_1^{\prime}(t)\Omega_N^{\prime}(t)-\Omega_1^{\prime}(t)\dot{\Omega}_N^{\prime}(t)]/\Omega^{\prime2}$.

Inspired by the above idea in section \textrm{IV}, we make the
system into a non-resonant system to construct the CDD Hamiltonian
in Eq. (33). Therefore, the Hamiltonian of the present system reads
$H_{total}^\prime=H_l^\prime+H_c^\prime+H_d^\prime$, where
$H_d^\prime=\sum_{i=1}^N\Delta|e\rangle\langle e|$. Similar to the
above procedure from Eq. (\ref{eq22}) to Eq. (\ref{eq23}) in Section
IV, we obtain the final effective Hamiltonian
\begin{eqnarray}
H_{fe(N)}^{\prime}=-\frac{\overline{\Omega}_1^{\prime2}}{3\Delta}|\phi_1^{\prime}\rangle\langle \phi_1^{\prime}|
-\frac{\overline{\Omega}_N^{\prime2}}{3\Delta}|\phi_{4N-1}^{\prime}\rangle\langle\phi_{4N-1}^{\prime}|
-\frac{\overline{\Omega}_1^\prime\overline{\Omega}_N^\prime}{3\Delta}(|\phi_1^{\prime}\rangle\langle\phi_{4N-1}^{\prime}|
+|\phi_{4N-1}^{\prime}|\rangle\langle\phi_1^{\prime}|).
\end{eqnarray}
For simplicity, we set $\overline{\Omega}_1^\prime=\overline{\Omega}_N^\prime=\overline{\Omega}^\prime$, the front two terms
caused by Stark shift can be omitted and the Hamiltonian becomes
\begin{eqnarray}
\overline{H}_N=\Omega_x^\prime(t)|\phi_1^\prime\rangle\langle\phi_{4N-1}^\prime|+H.c.,
\end{eqnarray}
where $\Omega_x^\prime(t)=-\frac{\overline{\Omega}^{\prime2}}{3\Delta}$.
To guarantee their consistency,
we put a change that $\Omega_N\rightarrow-i\Omega_N$. Then the eigenstates of $H_{eff(N)}$ become
\begin{eqnarray}
|\chi_o^\prime(t)\rangle=\left(
\begin{array}{c}
      \cos\theta^\prime(t) \\
      0\\
      i\sin\theta^\prime(t)
\end{array}
\right),
~|\chi_{\pm}^\prime(t)\rangle=\frac{1}{\sqrt{2}}\left(
\begin{array}{c}
      \sin\theta^\prime(t) \\
      \pm1\\
      -i\cos\theta^\prime(t)
\end{array}
\right),
\end{eqnarray}
and the CDD Hamiltonian $H(t)$ becomes
\begin{eqnarray}
H^\prime(t)=-\dot{\theta}^\prime|\phi_1^\prime\rangle\langle\phi_{4N-1}^\prime|-
\dot{\theta}^\prime|\phi_{4N-1}^\prime\rangle\langle\phi_1^\prime|.
\end{eqnarray}
Compared Eq. (35) with Eq. (37), we can easily get the CDD Hamiltonian
when the condition $\Omega_x^\prime=-\dot{\theta}^\prime$ is satisfied.
\begin{eqnarray}
\overline{\Omega}_1^\prime(t)=\overline{\Omega}_N^\prime(t)
=\overline{\Omega}^\prime(t)=\sqrt{3\Delta\dot{\theta}^\prime}.
\end{eqnarray}

\section{experimental feasibility and conclusions}
Now experimental feasibility needs to be discussed. The
configuration of $^{87}Rb$ can be suitable for our proposals. Under
current experimental condition a set of CQED parameters
$g=2\pi\times750MHz$, $\gamma=2\pi\times2.62MHz$, and
$\kappa_c=2\pi\times3.5MHz$ are available with the wavelength in the
region $630-850$nm \cite{SMSTJKKJVKWGEWHJKPRA}. By using fiber-taper
coupling to high-Q silica microspheres the efficiency of
fiber-cavity coupling is higher than $99.9\%$
\cite{SMSTJKOJPKJVPRL}. The optical fiber decay at a 852nm
wavelength is about 2.2dB/km \cite{KJGVFPDTGSB}, which means the
fiber decay rate is about $\kappa_f=1.52\times10^5Hz$. With the
above parameters, we obtain a relatively high fidelity about
$97.15\%$.

In conclusion, we have proposed an efficient scheme to fast
deterministically generate $N$-atom GHZ state in separate coupled
cavities via transitionless quantum driving (TQD) only by one-step
manipulation. We apply a promising method to construct STAP by joint
utilization of the Zeno dynamics and the approach of TQD in the
cavities QED system. The method features are that we do not need to
control the time exactly and the evolution process is fast. Because
the atoms are trapped in separate coupled cavity, the single qubit
manipulation can be realized easily. When considering dissipation,
we can see that the method is robust against the decoherences caused
by the atomic spontaneous emission and fiber decay. The results show
that the scheme has a high fidelity and may be possible to implement
with the current experimental technology. So, the scheme is fast,
robust and effective. We hope the scheme can be used to generate
multi-atom GHZ state in the future.

\section*{ACKNOWLEDGEMENT}

  This work was supported by the National Natural Science Foundation of China under Grants No.
11105030 and No. 11374054, the Foundation of Ministry of Education
of China under Grant No. 212085, and the Major State Basic Research
Development Program of China under Grant No. 2012CB921601.

\newpage

\begin{figure}
\scalebox{0.6}{\includegraphics{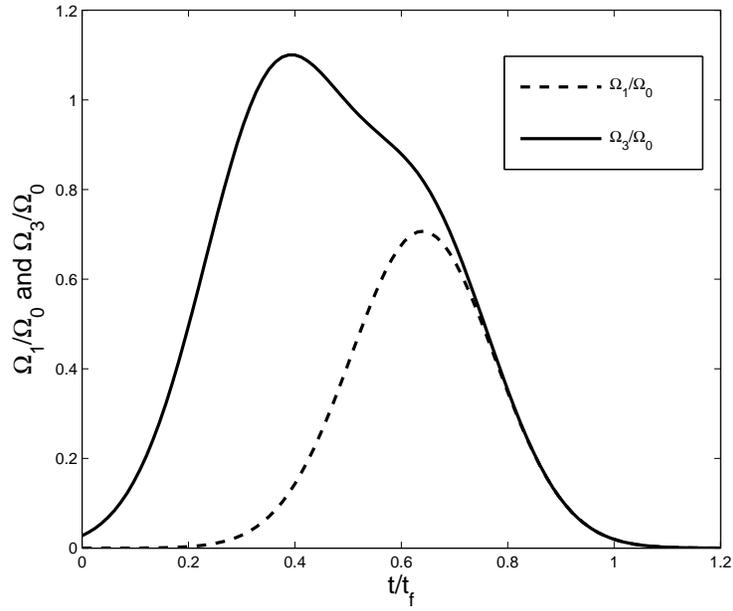}}
\caption{The laser pulses $\Omega_1/\Omega_0$ and $\Omega_3/\Omega_0$ versus $t/t_f$.}\label{FIG2}
\end{figure}

\begin{figure}
\scalebox{1.0}{\includegraphics{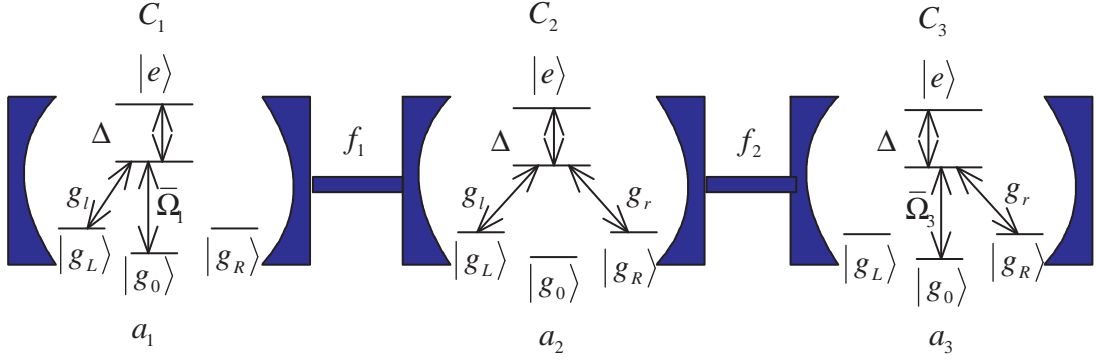}} \caption{The structure of
the experimental setup and atoms for the APF
Hamiltonian.}\label{FIG3}
\end{figure}

\begin{figure}
\scalebox{0.6}{\includegraphics{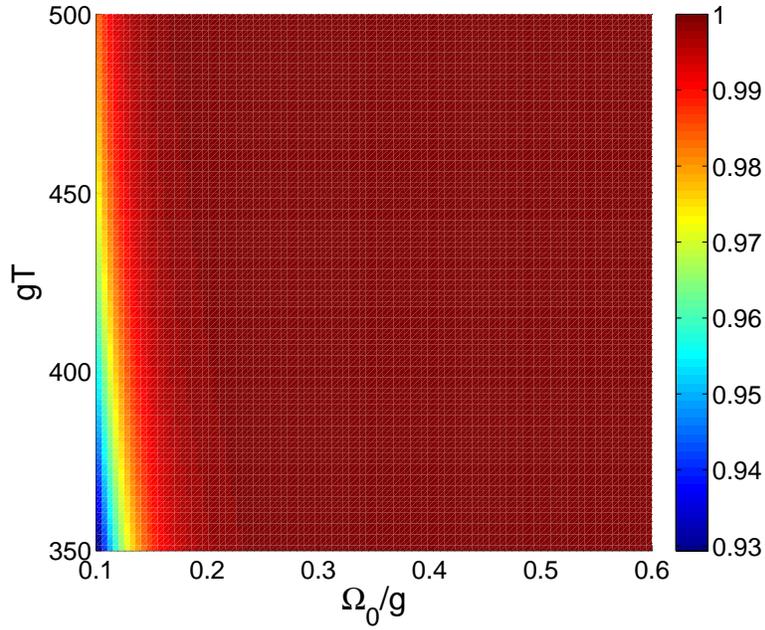}}
\caption{The fidelity versus the laser pulses amplitude $\Omega_0$ and the operation time $t/t_f$.}\label{FIG4}
\end{figure}

\begin{figure}
\renewcommand\figurename{\small FIG.}
\centering \vspace*{8pt}\setlength{\baselineskip}{10pt}
\subfigure[]{
\includegraphics[scale=0.4]{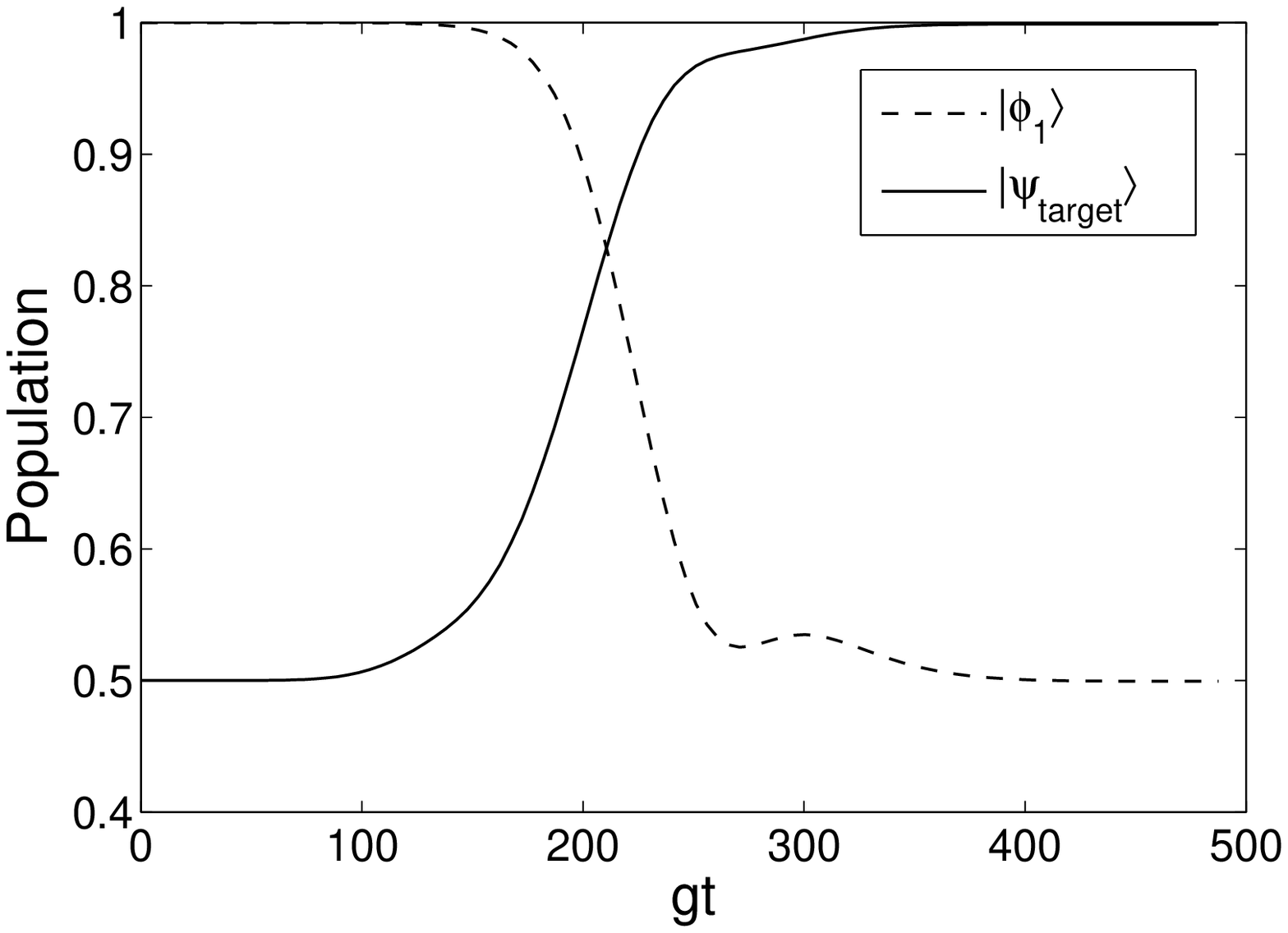}}
\subfigure[]{
\includegraphics[scale=0.4]{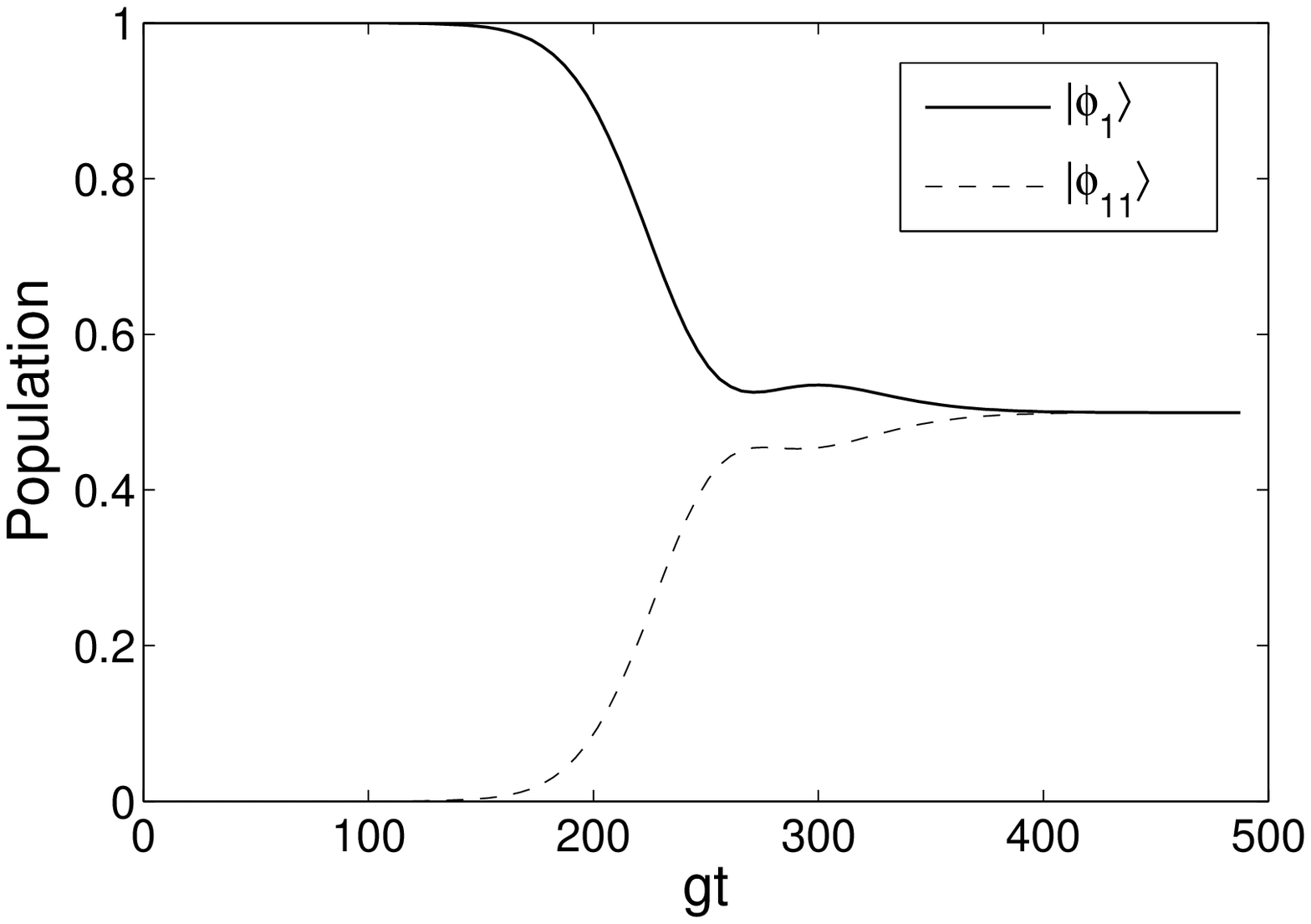}}
\caption{(a) The population $P_{target}$ of the target state $|\psi_{target}\rangle$ and the population $P_0$
of the initial state $|\psi(0)\rangle$ governed by the original Hamiltonian $H_{tot}$
via the adiabatic passage. (b) The population $P_1(t)$ of the states $|\phi_1\rangle$ and
the population $P_{11}(t)$ of the states $|\phi_{11}\rangle$ governed by the original
Hamiltonian $H_{tot}$ via adiabatic passage.
The parameters are collectively with the fixed values $\Omega_0=0.2g$, $g=v$, $t_0=0.14t_f$, $t_c=0.19t_f$ and
$t_f=400/g$.}\label{FIG5}
\end{figure}

\begin{figure}
\scalebox{0.6}{\includegraphics{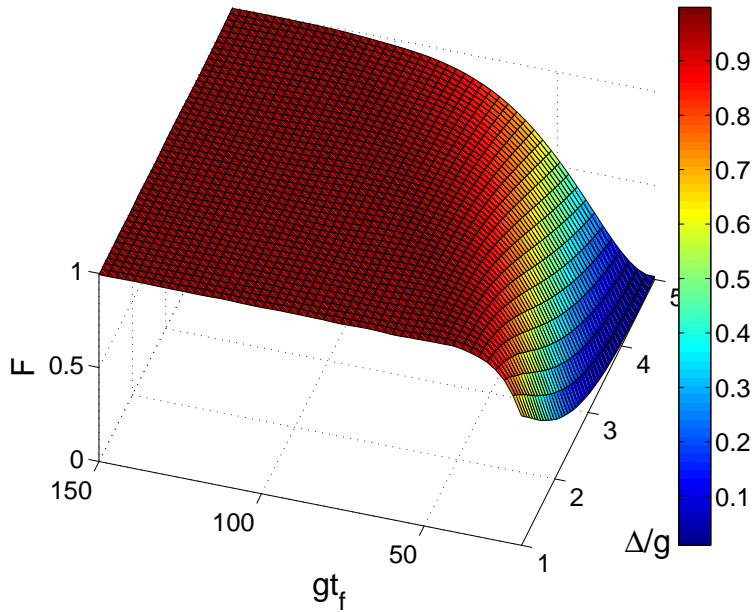}}
\caption{The fidelity $F$ of the target state $|\psi(t_f)\rangle$ governed by $H_{tot}^\prime$
versus the interaction time $gt_f$ and the detuning $\Delta/g$.}\label{FIG6}
\end{figure}

\begin{figure}
\renewcommand\figurename{\small FIG.}
\centering \vspace*{8pt}\setlength{\baselineskip}{10pt}
\subfigure[]{
\includegraphics[scale=0.4]{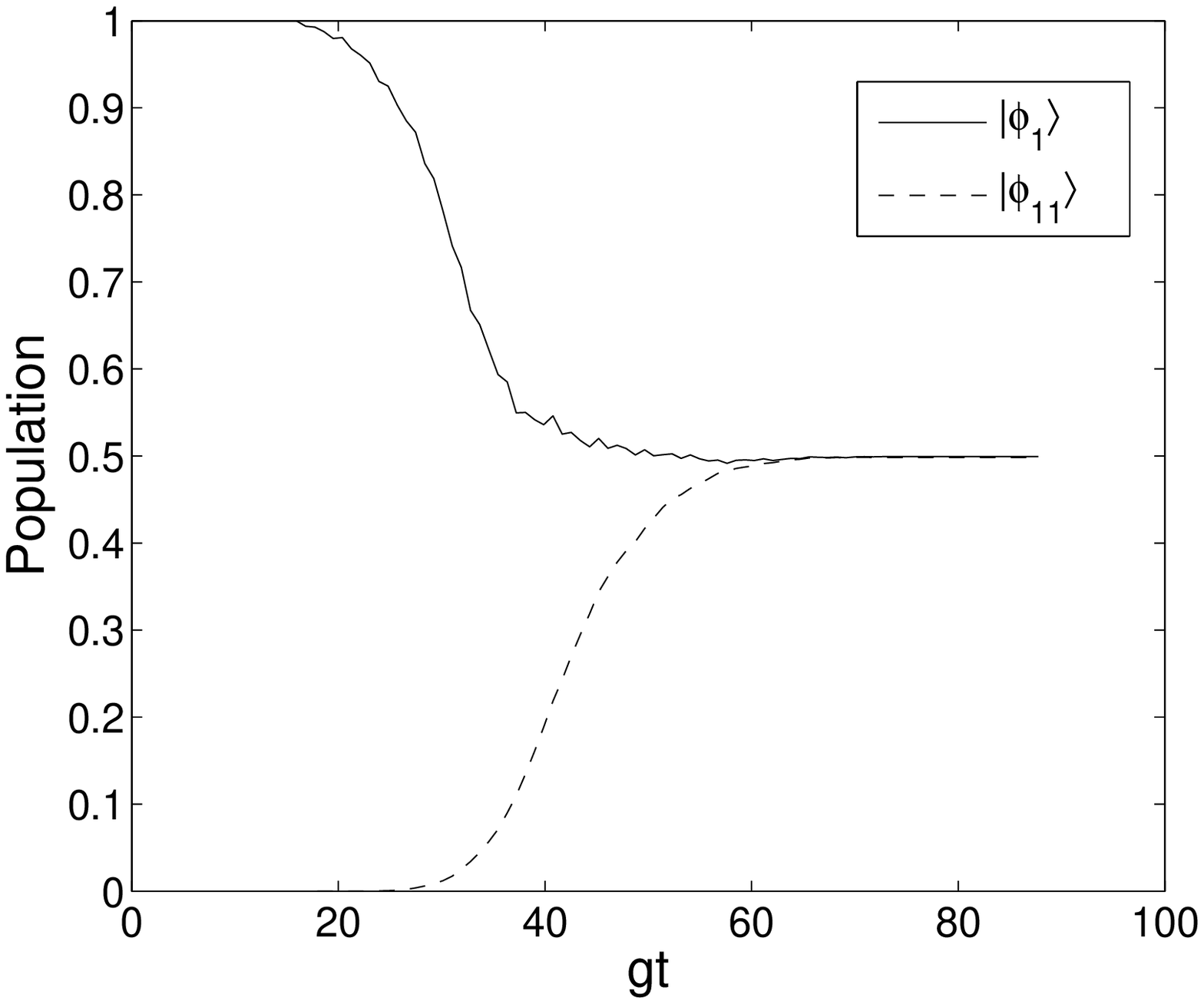}}
\subfigure[]{
\includegraphics[scale=0.4]{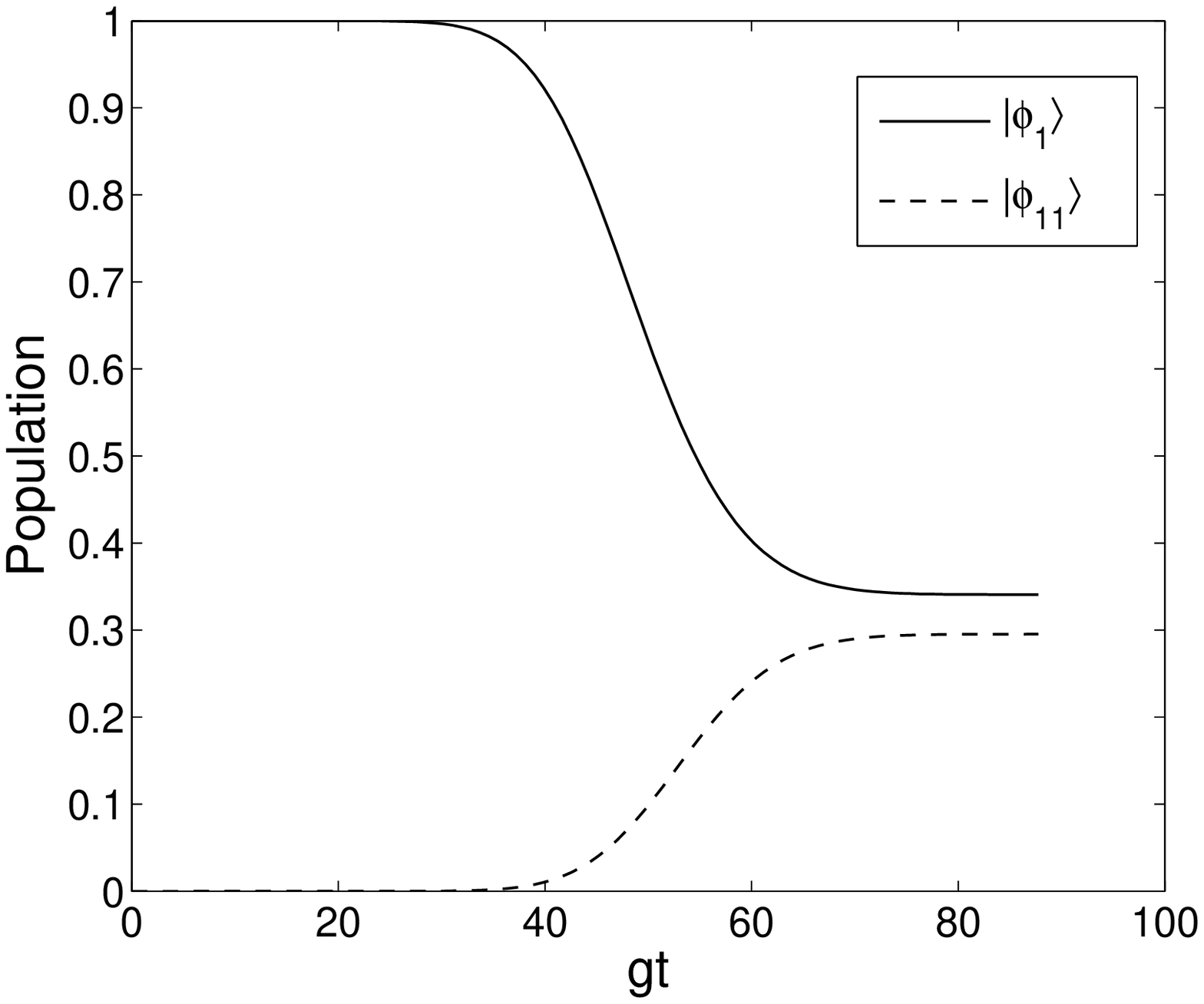}}
\caption{The population $P_1(t)$ of the state $|\phi_1\rangle$ and
the population $P_{11}(t)$ of the state $|\phi_{11}\rangle$ governed
by (a) the APF Hamiltonian $H_{tot}^\prime$ with $\Delta=2.3g$. (b)
The original Hamiltonian $H_{tot}$ collectively with the fixed
values $\Omega_0=0.2g$, $g=v$, $t_0=0.14t_f$, $t_c=0.19t_f$, and
$t_f=72/g$.}\label{FIG7}
\end{figure}

\begin{figure}
\renewcommand\figurename{\small FIG.}
\centering \vspace*{8pt}\setlength{\baselineskip}{10pt}
\subfigure[]{
\includegraphics[scale=0.42]{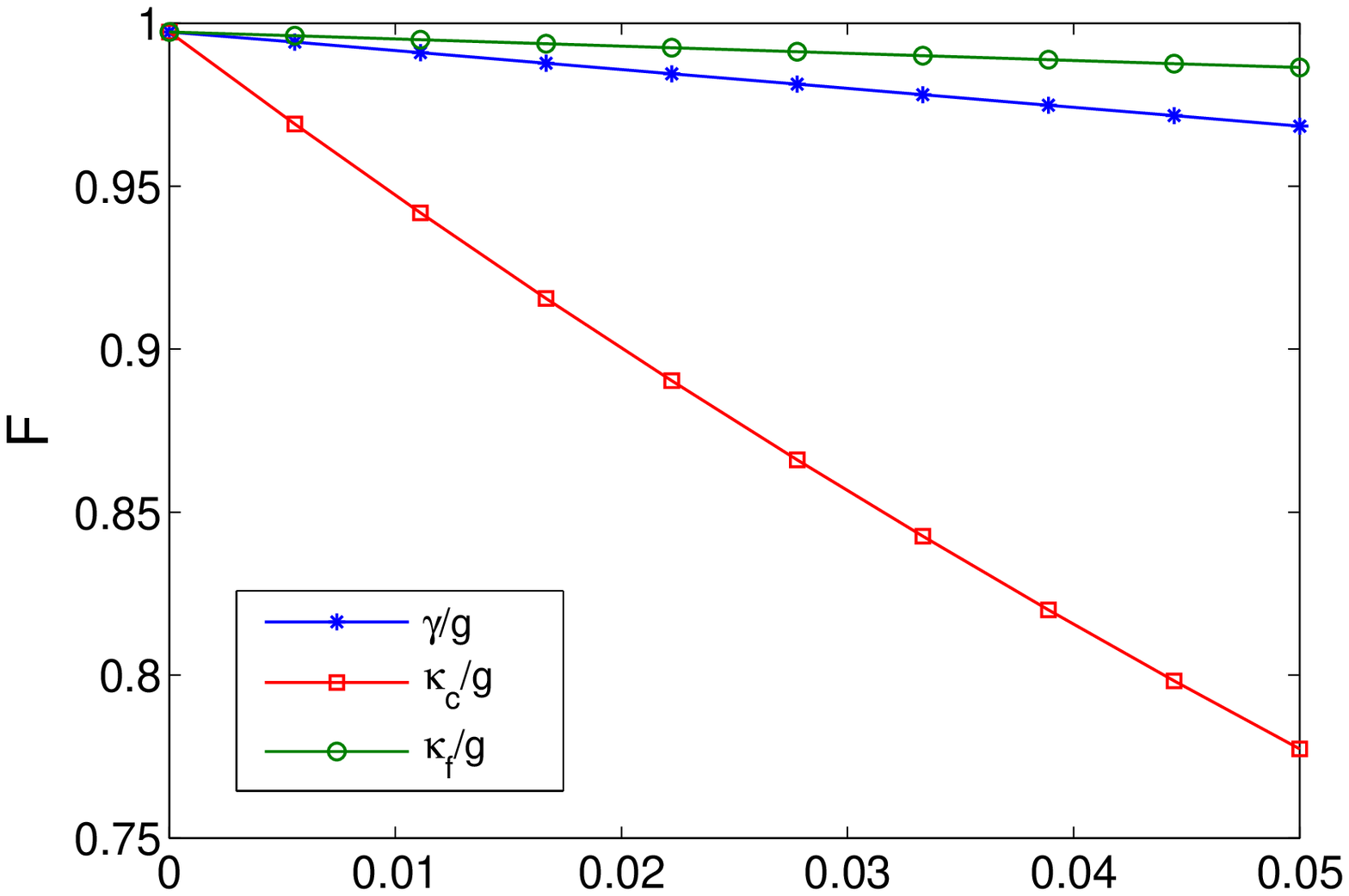}}
\subfigure[]{
\includegraphics[scale=0.42]{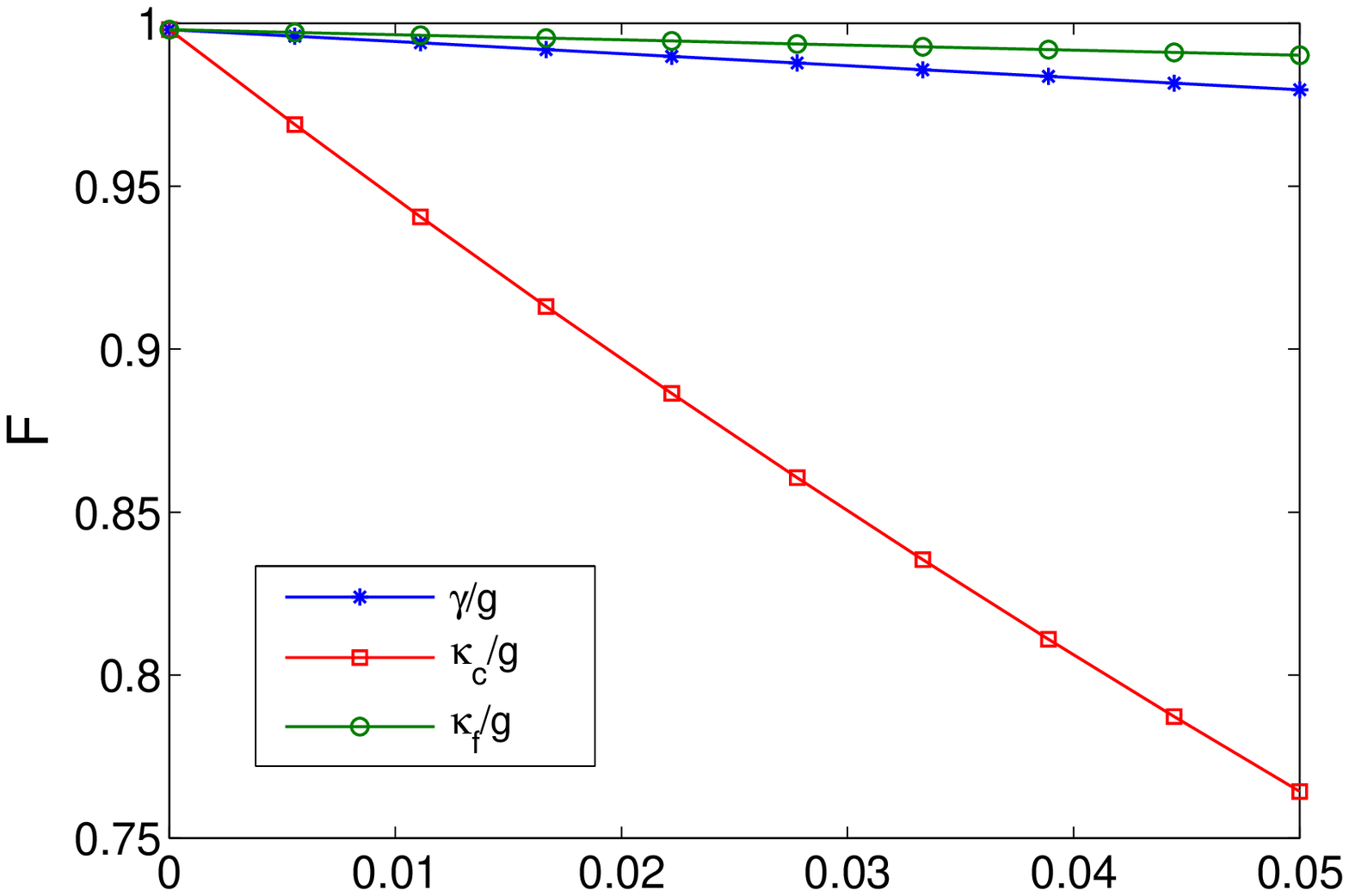}}
\caption{The fidelity of the target state $|\psi(t_f)\rangle$
 governed by (a) the APF Hamiltonian $H_{tot}^\prime$ with
$\Delta=2.3g$, $t_f=72/g$ and $\Omega_0=0.2g$. (b) the original
Hamiltonian $H_{tot}$ with $t_f=153/g$, and $\Omega_0=0.5g$
collectively with the fixed values $t_0=0.14t_f$, and $t_c=0.19t_f$
versus the dimensionless parameters $\gamma/g$, $\kappa_c/g$, and
$\kappa_f/g$, respectively.}\label{FIG8}
\end{figure}

\begin{figure}
\renewcommand\figurename{\small FIG.}
\centering \vspace*{8pt}\setlength{\baselineskip}{10pt}
\subfigure[]{
\includegraphics[scale=0.4]{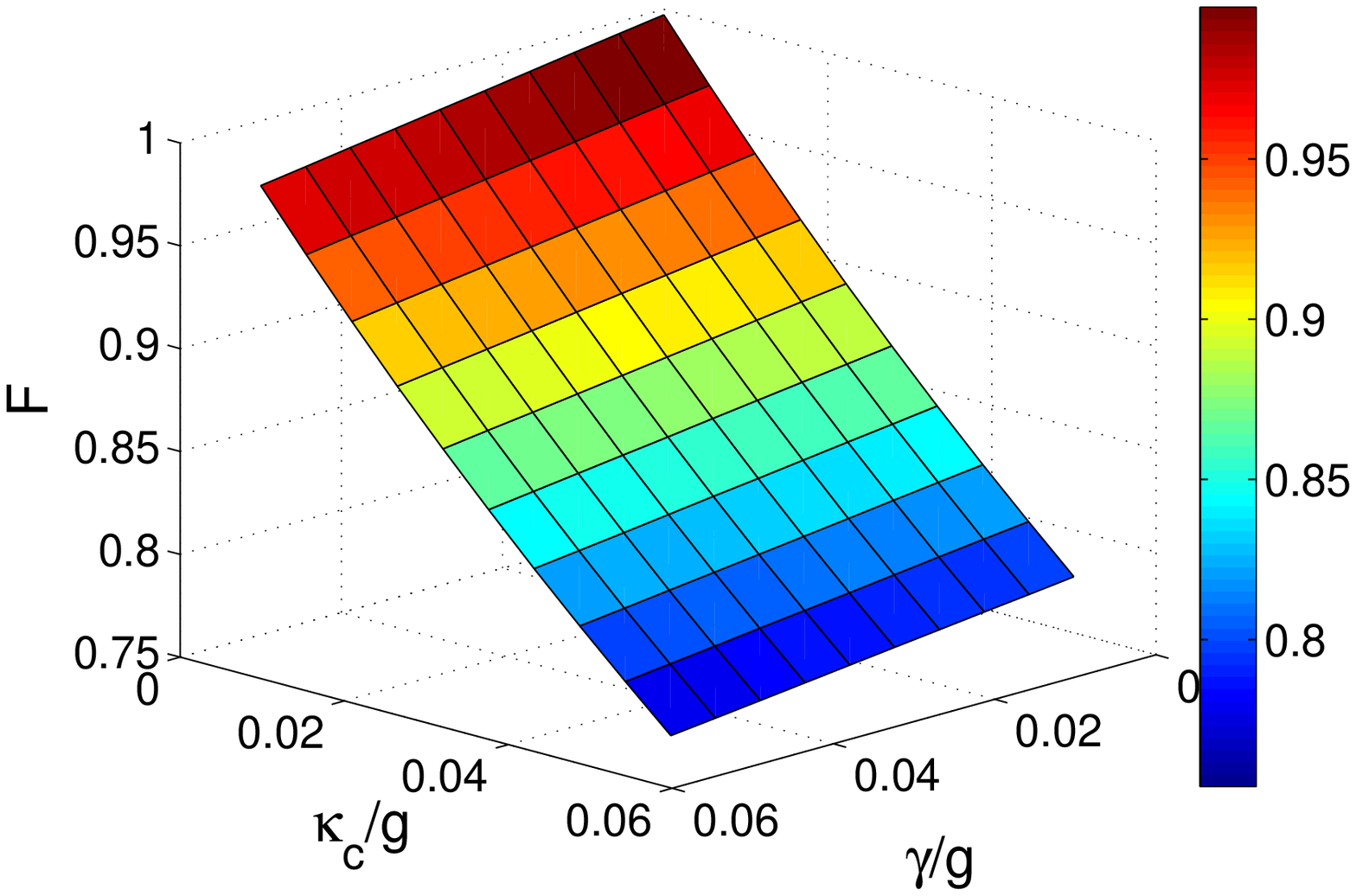}}
\subfigure[]{
\includegraphics[scale=0.4]{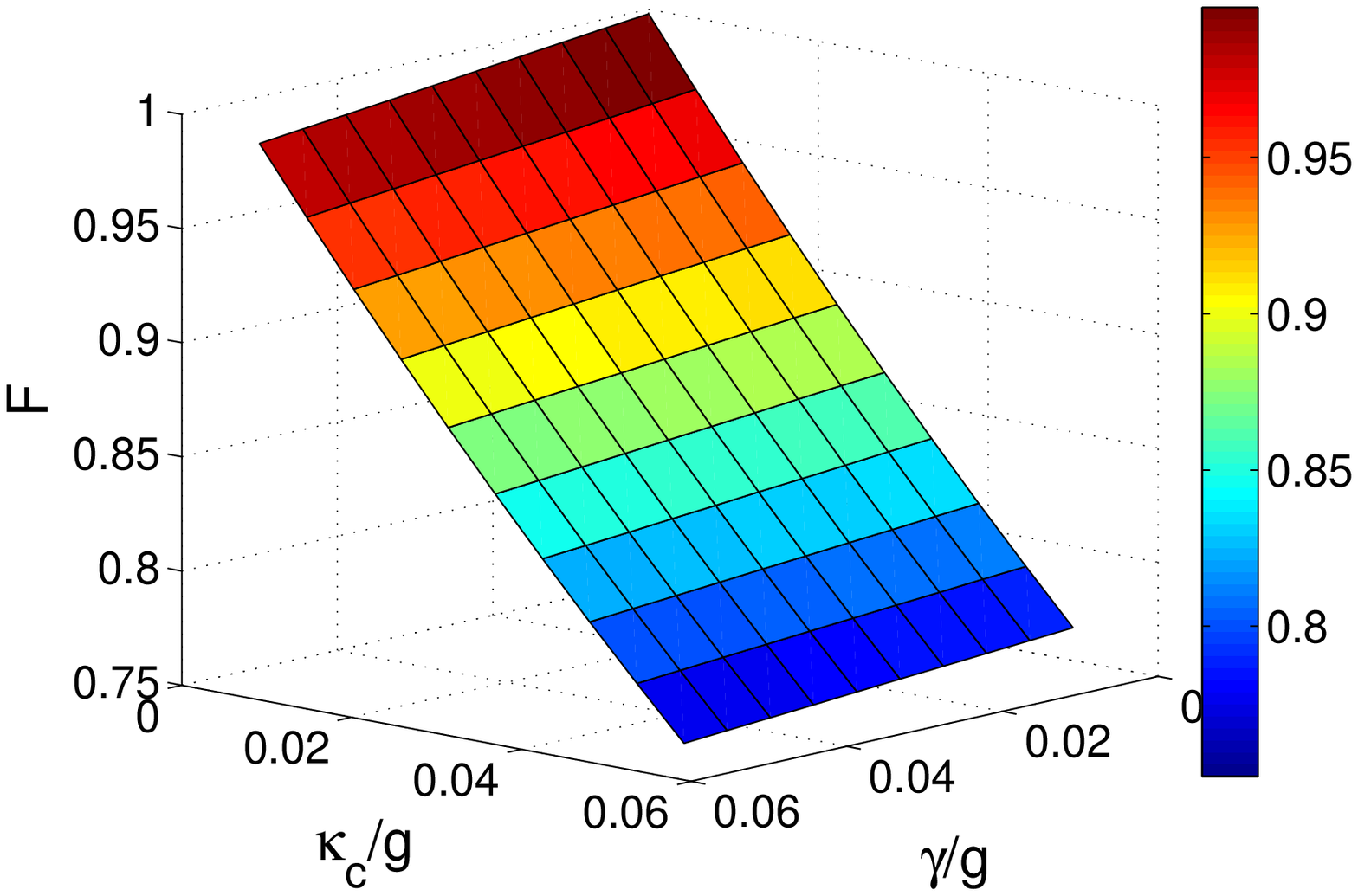}}
\caption{The fidelity of the target state $|\psi(t_f)\rangle$
 governed by (a) the APF Hamiltonian $H_{tot}^\prime$ with
$\Delta=2.3g$, $t_f=72/g$, and $\Omega_0=0.2g$. (b) the original
Hamiltonian $H_{tot}$ with $t_f=153/g$, and $\Omega_0=0.5g$
collectively with the fixed values $t_0=0.14t_f$, and $t_c=0.19t_f$
versus the dimensionless parameters $\gamma/g$ and
$\kappa_c/g$.}\label{FIG9}
\end{figure}

\begin{figure}
\renewcommand\figurename{\small FIG.}
\centering \vspace*{8pt}\setlength{\baselineskip}{10pt}
\subfigure[]{
\includegraphics[scale=0.35]{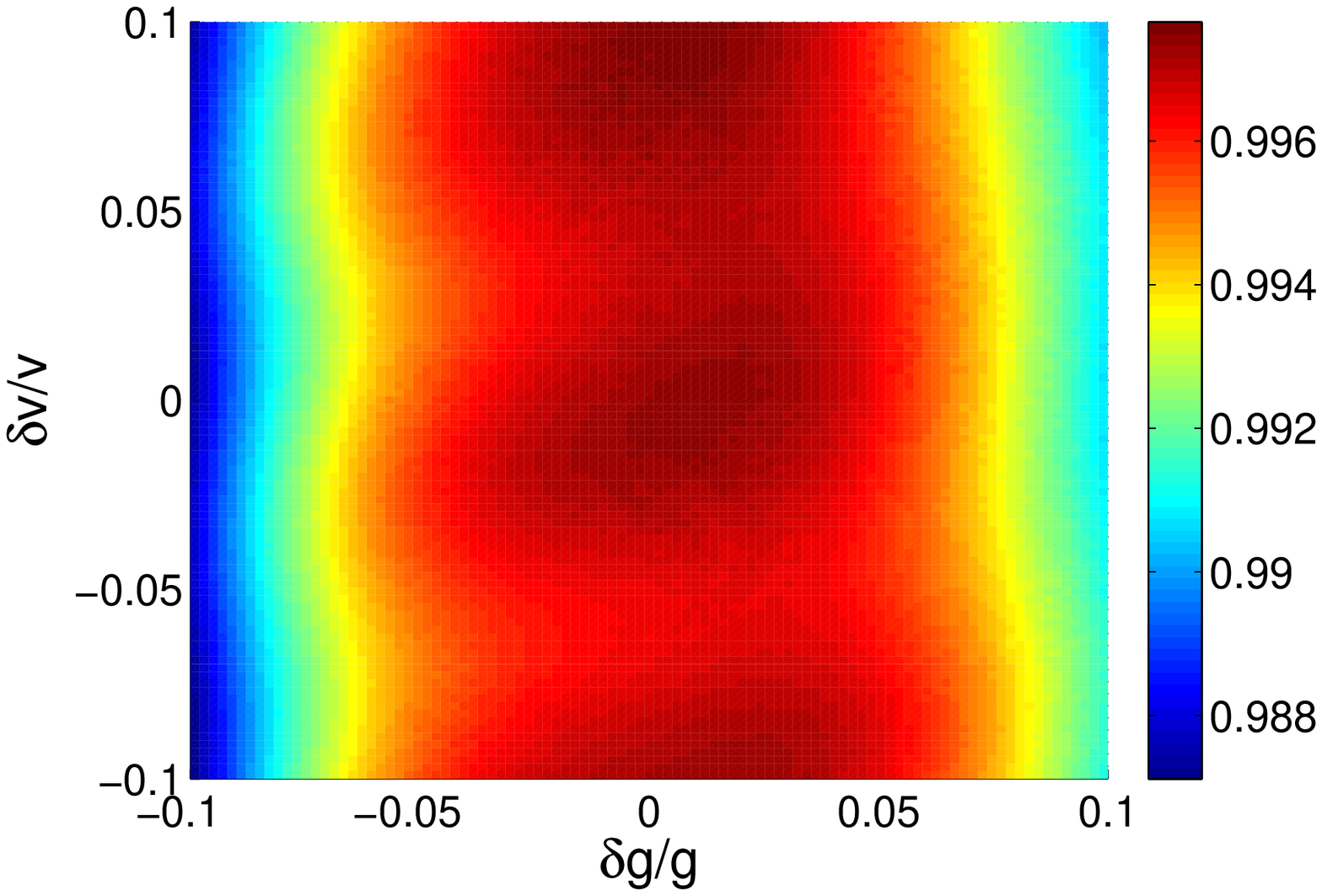}}
\subfigure[]{
\includegraphics[scale=0.35]{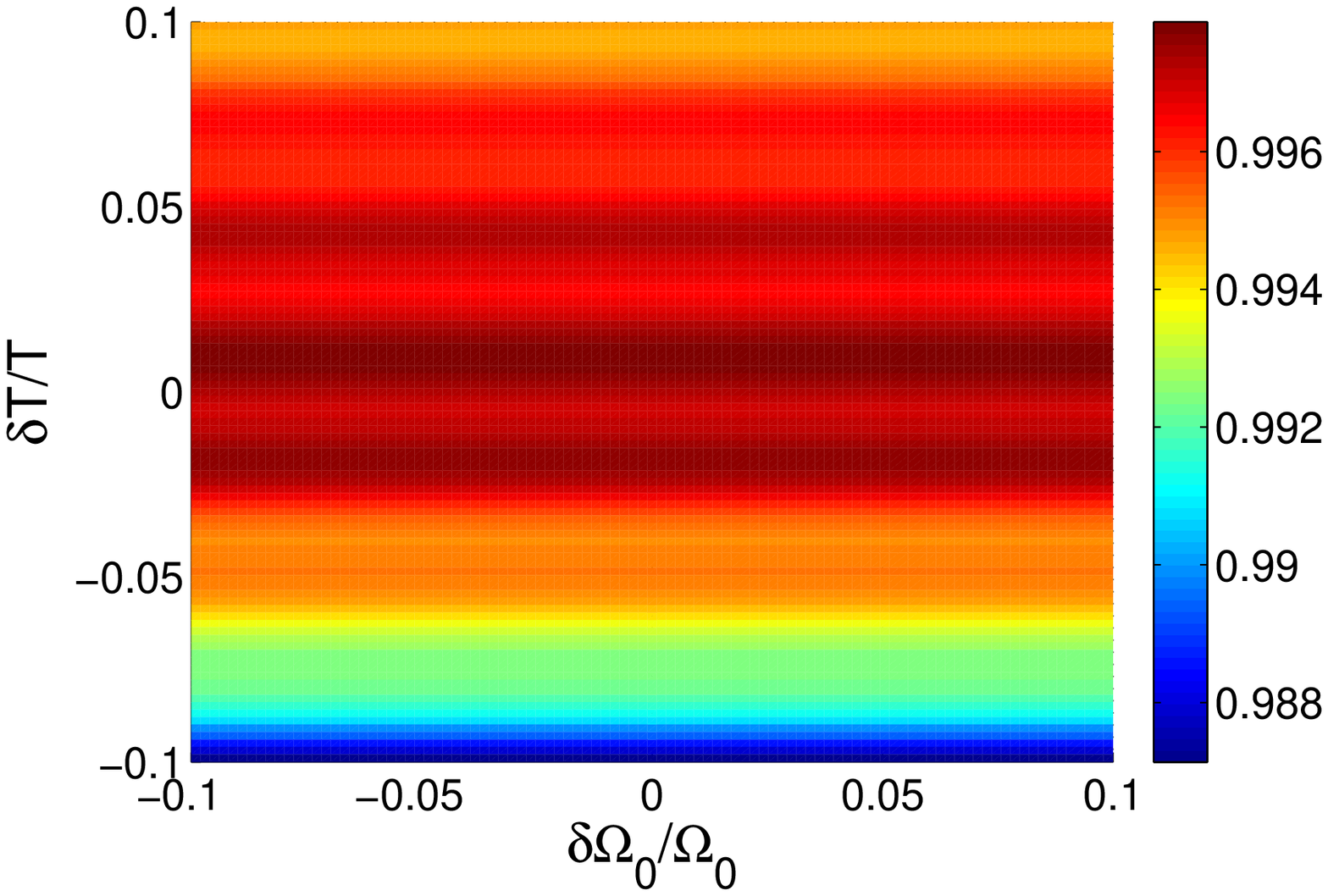}}
\caption{The fidelity of the target state $|\psi_{target}\rangle$
versus the deviations of (a) $g$ and $v$, (b) $T$ and
$\Omega_0$.}\label{FIG10}
\end{figure}

\begin{figure}
\scalebox{1.0}{\includegraphics{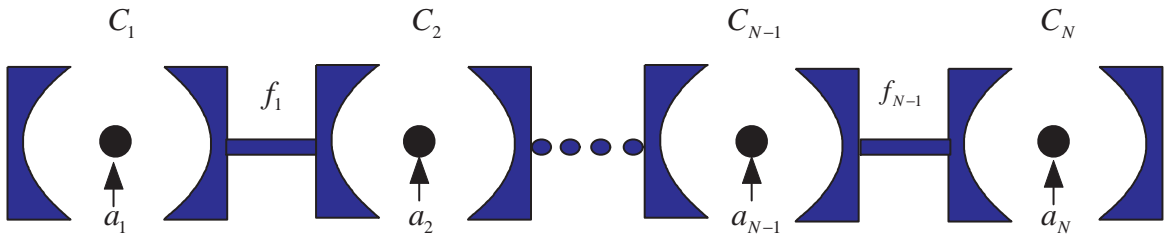}} \caption{The set-up
diagram for the generation of $N$-atom GHZ states. The $N$-atoms are
respectively trapped in $N$-cavities which are linked by $N-1$
fibers.}\label{FIG11}
\end{figure}

\end{document}